 \documentclass[pmlr,twocolumn,10pt]{jmlr} 

\usepackage{booktabs}
\usepackage{siunitx}

\usepackage{bm}

\usepackage{tikz}
\usepackage{pgfplots}
\pgfplotsset{compat=1.18}
\usepackage[switch]{lineno}
\usepackage{enumitem}
\usepackage[utf8]{inputenc}
\usepackage{booktabs, threeparttable, makecell, xcolor}
\usepackage{relsize} 
\definecolor{lightgray}{gray}{0.8}

\usepackage{titlesec}

\titlespacing*{\section}
  {0pt}   
  {10pt} 
  {5pt} 

\theorembodyfont{\upshape}
\theoremheaderfont{\scshape}
\theorempostheader{:}
\theoremsep{\newline}

\jmlrvolume{297}
\jmlryear{2025}
\jmlrworkshop{Machine Learning for Health (ML4H) 2025} 

 \title[Beyond the Clinic: Augmenting EHR with Wearable Data for Health Prediction]{Beyond the Clinic: A Large-Scale Evaluation of Augmenting EHR with Wearable Data for Diverse Health Prediction}

\author{%
\Name{Will Ke Wang} \Email{kw3215@cumc.columbia.edu}\\
\addr Columbia University
\AND
\Name{Rui Yang} \Email{yang\_rui@u.nus.edu}\\
\addr National University of Singapore
\AND
\Name{Chao Pang} \Email{cp3016@cumc.columbia.edu}\\
\addr Columbia University
\AND
\Name{Karthik Natarajan} \Email{kn2174@cumc.columbia.edu}\\
\addr Columbia University
\AND
\Name{Nan Liu} \Email{liu.nan@duke-nus.edu.sg}\\
\addr National University of Singapore
\AND
\Name{Daniel McDuff} \Email{dmcduff@google.com}\\
\addr Google
\AND
\Name{David J Slotwiner} \Email{djs2001@med.cornell.edu}\\
\addr NewYork-Presbyterian in Queens
\AND
\Name{Fei Wang} \Email{few2001@med.cornell.edu}\\
\addr Weill Cornell Medicine
\AND
\Name{Matthew B.A. McDermott} \Email{mm6677@cumc.columbia.edu}\\
\addr Columbia University
\AND
\Name{Xuhai Xu} \Email{xx2489@cumc.columbia.edu}\\
\addr Columbia University
}

\begin{document}

\maketitle
\begin{abstract}
Electronic health records (EHRs) provide a powerful basis for predicting the onset of health outcomes. Yet EHRs primarily capture in-clinic events and miss aspects of daily behavior and lifestyle containing rich health information. Consumer wearables, by contrast, continuously measure activity, heart rate, and sleep, and more, offering complementary signals that can fill this gap. Despite this potential, there has been little systematic evaluation of the benefit that wearable data can bring to health outcome prediction on top of EHRs. In this study, we present an extensible framework for multimodal health outcome prediction that integrates EHR and wearable data streams. Using data from the All of Us Program, we systematically compared the combination of different encoding methods on EHR and wearable data, including the traditional feature engineering approach, as well as foundation model embeddings. Across ten clinical outcomes, wearable integration consistently improved model performance relative to EHR-only baselines, e.g., average $\Delta$AUROC +6.8\% for major depressive disorder, +9.7\% for hypertension, and +12.6\% for diabetes. On average across all ten outcomes, fusing EHRs with wearable features shows 8.5\% improvement in AUROC.
To our knowledge, this is the first large-scale evaluation of wearable–EHR fusion, underscoring the utility of wearable-derived signals in complementing EHRs and enabling more holistic, personalized health outcome predictions.
Meanwhile, our analysis elucidates future directions for optimizing foundation models for wearable data and its integration with EHR data.
\end{abstract}

\paragraph*{Data and Code Availability}
Data used for this study is available upon approval through the \href{https://workbench.researchallofus.org/login}{\textit{All of Us} Research Program}. The codebase is made available at \href{https://github.com/WillKeWang/AoU_Wearable_Valuation.git}{GITHUB/AoU\_Wearable}.

\paragraph*{Institutional Review Board (IRB)} Redacted

\section{Introduction}
\label{sec:intro}
Electronic health records (EHRs) have become an important data foundation for predicting health outcomes in the past decades, with predictive models developed from them demonstrating strong performance across a variety of clinical tasks~\citep{rajkomar_scalable_2018, goldstein_opportunities_2017}. However, EHRs only capture episodic, clinic-centered snapshots of health status \citep{goldstein_opportunities_2017}, missing the continuous physiological and behavioral patterns unfolding in patients' daily lives between clinical encounters. This temporal sparsity represents a critical blind spot: while EHRs excel at documenting what happens during healthcare interactions, they cannot capture daily health indicators such as deviations in resting heart rate, changes in activity patterns, or sleep disruptions that often precede clinical events and could enable earlier intervention \citep{liu_digital_2025,xu_globem_2023}.

\begin{figure*}[htbp]
    \vspace{-8mm}
  \includegraphics[width=1.0\linewidth]{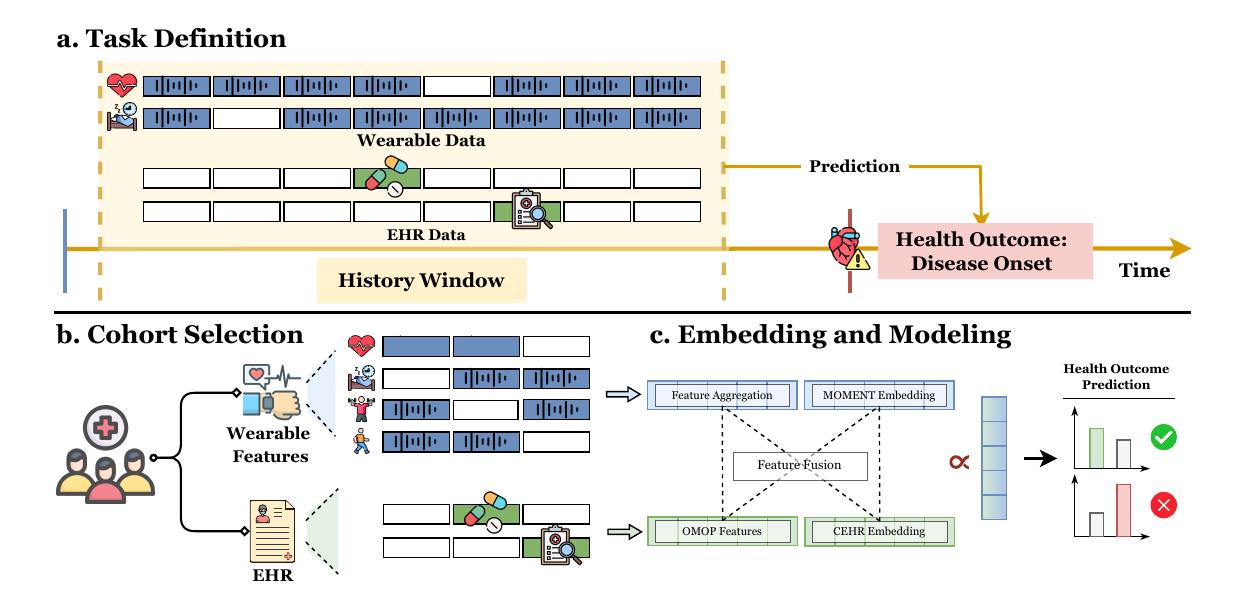}
  \label{fig:overview}
  \vspace{-8mm}
  \caption{Overall workflow. (a) Task definition: we use historical EHR and wearable data to predict for disease onset recorded in the EHR system (b) Cohort Definition: Participants with both EHR and wearable data were selected and required to meet data quality control criteria. (c) Embedding and Modeling: Wearable data were encoded using feature engineering and MOMENT embeddings, while EHR data were represented using OMOP indicators and CEHR embeddings. Three fusion strategies—concatenation, weighted concatenation, and feature selection—were applied, and the fused representations were used for health outcome prediction.}
  \vspace{-5mm}
\end{figure*}

Consumer wearable devices offer a solution to address this gap, providing continuous behavrioral and physiological data streams to capture everyday behavior patterns that embed health-related information complementary to EHRs~\citep{li_digital_2017, null_all_2019,ginsburg_key_2024} and have the potential to enhance EHR-based predictive models. 
Recent studies have shown that at the population scale, daily behaviors captured by consumer wearables link to various clinical outcomes:~\citet{master_association_2022} ``identified consistent and statistically significant associations between activity levels and incident diabetes, hypertension, GERD, MDD, obesity and sleep apnea'';
~\citet{zheng_sleep_2024} has found that irregular timing and shorter sleep relate to higher hazards for conditions including hypertension, MDD, and generalized anxiety disorder (GAD).
Recent work has started to evaluate ML models on wearable data to predict health outcomes.
For example, \citet{kundrick_machine_2025} built wearable-based models to predict future hospitalization and incident cardiovascular disease (CVD).
\citet{xu_globem_2023} built algorithms on wearable data to predict young adults' depressive symptoms in daily behavior.

Building on this encouraging signal, there are emerging efforts that combine EHRs and wearable data for clinical outcome prediction. 
\citet{modde_epstein_linking_2023} conducted a feasibility study linking EHRs with Fitbit data and showed that longitudinal heart rate patterns during pregnancy reveal distinct physiological changes and could help detect maternal health problems early.
Some studies combined EHR data with wearable sensor features to predict hospital readmission, demonstrating that the combined signals can significantly improve model performance ~\citep{yhdego_prediction_2023,nagarajan2024impact}. However, most of these works are either limited by small, single-center cohorts or focused on narrowly scoped clinical questions of particular health outcomes. The value of wearable data to enhance EHR-based predictive models has not been validated in any systematic, large-scale setup.

In this work, \textbf{we conduct the first comprehensive and systematic evaluation of EHR and wearable integration for multimodal predictive models across diverse disease outcomes} by leveraging the recent \textit{All of Us} Program that offers large cohorts with longitudinal wearable data (Fitbit) linked with their EHRs (formatted in Observational Medical Outcomes Partnership OMOP Common Data Model)~\citep{null_all_2019}.

We build a standardized, reproducible data processing and modeling pipeline to fuse EHR and wearable modalities.
We systematically evaluate our pipeline across ten diseases:
Hyperlipidemia, Hypertension, Gastroesophageal Reflux Disease (GERD), Generalized Anxiety Disorder (GAD), Major Depressive Disorder (MDD), Obesity, Sleep Apnea, Type II Diabetes, Heart Failure, Atrial Fibrillation.
Figure~\ref{fig:overview} presents the task setup.
A common challenge in this effort is determining the most effective way to represent and integrate these distinct data modalities. To address this, we draw upon recent advancements in representation learning to compare a traditional feature engineering approach with a state-of-the-art foundation model for each data type.
For EHR data, we evaluate both the traditional method of high-dimensional OMOP indicators and a sequence-generative approach, CEHR-GPT, which preserves chronological patient timelines and enables reproducible, shareable representations~\citep{pang_cehr-gpt_2024}. 
For wearable data, we explore both a manual feature engineering approach that aggregates behavioral statistics over time and a time-series foundation model, MOMENT, which can provide general-purpose wearable data embeddings~\citep{goswami_moment_2024}.
Finally, we evaluate three strategies for fusing these varied representations: concatenation, adjusted weighting, and top-k feature selection.

Our results show that integrating wearable data yields consistent and substantial performance gains for most conditions regardless of the encoding and fusion approaches. \textbf{On average across all ten outcomes, fusing EHRs with wearable features shows 8.5\% improvement in Area under the ROC curve (AUROC).}
Beyond these results, our comparisons of encoding methods offer deeper insights into the limitations of current general-purpose foundation models when applied to health outcome prediction. Our contributions are three-fold: 
\begin{itemize}[itemsep=0pt, parsep=0pt]
    \item We present the first large-scale, systematic benchmark on integrating consumer wearable data with EHRs for predicting ten diverse health outcomes. Our work provides a comprehensive and nuanced analysis of the significant predictive value that wearable data adds, quantifying the magnitude of improvement across different health conditions.
    \item Our analyses further identify the limitations of the state-of-the-art foundational models in health outcome prediction tasks, which shed light on future priority areas to advance EHR and wearable foundational models for AI health analytics.
    \item We open-source a standardized, reproducible, and extensible data processing and modeling pipeline to fuse EHR and wearable modalities for developing future multimodal health prediction models.
\end{itemize}

\section{Methods}
\label{sec:methods}


In this work, we address the predictive task of determining whether a patient will develop a given condition in the future based on their historical EHR data combined with continuous wearable sensor streams collected prior to diagnosis (pre-index period).
Our experimental pipeline consists of the following key components:
1) Cohort Definition (Sec.~\ref{sec:cohort}) - finding clean cohorts with available EHR and wearable data and distinct disease onset;
2) Feature Extraction (Sec.~\ref{sec:features}) - comparing traditional feature engineering against foundation model embeddings for both EHR and wearable modalities;
3) Multimodal Feature Fusion (Sec.~\ref{sec:fusion}) - evaluating strategies to optimally combine these distinct data types;
and 4) Systematic Evaluation (Sec.~\ref{sec:evaluation}) - benchmarking performance across ten diverse clinical outcomes using the All of Us dataset.
Our framework is extensible to any clinical outcome of interest and allows us to quantify the added predictive value of wearable data beyond EHR-only baselines while exploring new approaches for multimodal health prediction.

\subsection{Cohort Definition}
\label{sec:cohort}
We implemented a rigorous cohort definition to develop high-quality datasets for clinical outcome prediction using integrated EHR and wearable sensor data from the \textbf{All of Us registered tier}. For participants with both EHR and wearable data, we applied systematic wearable data filters to ensure reliable physiological measures. A valid day required $\geq$240 minutes of sleep, heart rate between 30–240 bpm, and $<$90-minute discrepancy between total sleep time and sum of stages to confirm valid sleep architecture. Wearable records were aligned by defining the first valid day as the first with both valid sleep and heart rate, and the last valid day accordingly. Participants were retained only if $\geq$50\% of days between these bounds were valid and the difference between the first and last valid days were at least 180 valid days.

A positive outcome was defined as evidence of the target disease via OMOP concept IDs, disease-specific prescriptions, or ICD-to-PheCode–mapped codes. To capture incident rather than pre-existing disease, we followed the practice of \citet{master_association_2022} and required the first evidence to occur $\geq$180 days after wearable tracking start, creating a temporal buffer to separate new-onset from prior conditions and ensuring sufficient wearable data.
In this work, we focus on 10 disease outcomes that are well-represented in the All of Us dataset, per suggestion by \citet{master_association_2022}.

All participants—positive and negative—needed healthcare engagement before wearable monitoring (to verify absence of prior disease indicators for positives and establish baseline activity) and after the observation period (to confirm continued engagement and follow-up). For incident cases, prediction time was anchored to one day before the first disease evidence (diagnostic PheCode, SNOMED, or drug code). For negatives, prediction time was one day before their last EHR record.

To ensure feature quality, we excluded participants with EHR data with invalid visit occurrence ids, or with insufficient or problematic sensor data. To prevent data leakage, we implemented train-test splitting that restricted held-out test sets to participants not present in CEHR model training data set. See Appendix A and Supplementary \figureref{fig:cohort_example} for detailed example using major depressive disorder as an exemplar outcome. See \appendixref{apd:demographics} for demographics break down for the final dataset used for each of the 10 outcomes.

\begin{table*}[t]
\centering
\footnotesize
\caption{Summary of fusion approaches combining EHR and wearable data representations.}
\label{tab:fusion_methods}
\resizebox{\textwidth}{!}{%
\begin{tabular}{l l l p{\dimexpr\textwidth-6.5cm-6\tabcolsep-2\arrayrulewidth\relax} l}
\toprule
\begin{tabular}[t]{@{}l@{}}\textbf{EHR}\\ \textbf{Source}\end{tabular} & \begin{tabular}[t]{@{}l@{}}\textbf{Wearable}\\ \textbf{Source}\end{tabular} & \begin{tabular}[t]{@{}l@{}}\textbf{Fusion}\\ \textbf{Strategy} \end{tabular} & \textbf{Feature Fusion Strategy} & \textbf{Overall Method} \\
\midrule
OMOP & \begin{tabular}[t]{@{}l@{}}Summary\\ Features\end{tabular} & \begin{tabular}[t]{@{}l@{}}Top-Features\\ Concatenation\end{tabular} & Univariate screening: top 50\% wearable features + equal number of OMOP indicators, then concatenation & [$\textbf{x}_{\text{EHR-OMOP}}$, $\textbf{x}_{\text{wear-sum}}$] (Concat) \\
\cmidrule{2-5}
& \begin{tabular}[t]{@{}l@{}}Time-Series\\ Embedding\end{tabular} & \begin{tabular}[t]{@{}l@{}}Top-Features\\ Concatenation\end{tabular} & Univariate screening: top OMOP indicators + top 50\% (count = 512) time-series dimensions, then concatenation & [$\textbf{x}_{\text{EHR-OMOP}}$, $\textbf{x}_{\text{wear-ts}}$] (Concat) \\
\midrule
CEHR-GPT & \begin{tabular}[t]{@{}l@{}}Summary\\ Features\end{tabular} & \begin{tabular}[t]{@{}l@{}}Concatenation \\ with Substitution\end{tabular} & Replace 221 least-informative CEHR dimensions with 221 wearable features based on linear classifier coefficients & [$\textbf{x}_{\text{EHR-CEHR}}$, $\textbf{x}_{\text{wear-sum}}$] (Concat) \\
\cmidrule{2-5}
& & \begin{tabular}[t]{@{}l@{}}Weighted\\ Concatenation\end{tabular} & Weighted concatenation $X_{\text{fused}} = [\alpha \textbf{x}_{\text{EHR-CEHR}} | \beta \textbf{x}_{\text{wear-sum}}]$ with $(\alpha,\beta)$ optimized on 70/30 dev split & [$\textbf{x}_{\text{EHR-CEHR}}$, $\textbf{x}_{\text{wear-sum}}$] (Weighted) \\
\cmidrule{2-5}
& \begin{tabular}[t]{@{}l@{}}Time-Series\\ Embedding\end{tabular} & \begin{tabular}[t]{@{}l@{}}Direct\\ Concatenation\end{tabular} & Direct concatenation of CEHR (768-dim) with time-series embedding (1024-dim) & [$\textbf{x}_{\text{EHR-CEHR}}$, $\textbf{x}_{\text{wear-ts}}$] (Concat) \\
\cmidrule{2-5}
& & \begin{tabular}[t]{@{}l@{}}Weighted\\ Concatenation\end{tabular} & Weighted concatenation with $(\alpha,\beta)$ in $[\alpha \textbf{x}_{\text{EHR-CEHR}} | \beta \textbf{x}_{\text{wear-sum}}]$ optimized on 70/30 dev split & [$\textbf{x}_{\text{EHR-CEHR}}$, $\textbf{x}_{\text{wear-ts}}$] (Weighted) \\
\bottomrule
\end{tabular}
}
\begin{tablenotes}[flushleft]\footnotesize
\item \textbf{Notes:} All features were z-score standardized on the training set prior to fusion. Univariate screening based on AUROC performance.
\end{tablenotes}
\vspace{-3mm}
\end{table*}

\vspace{-2mm}
\subsection{Feature Extraction}
\label{sec:features}
We employed two representation methods for EHR data and two methods for wearable data that are aligned to each participant’s pre-index window.
For EHR, we have (i) OMOP binary indicators, (ii) a structured-EHR foundation model embedding (CEHR-GPT).
For wearable data, we have (i) hand-engineered wearable summaries, and (ii) wearable time-series foundation embeddings (MOMENT).
All features were computed \emph{only} from data occurring before the index. All fusion variants applied the same scaler and feature-selection indices. 
\vspace{-3mm}
\subsubsection{\textbf{EHR Data Representation}}
\textbf{(i) OMOP binary indicators (\texttt{inspectOMOP})}

Using the \texttt{inspectOMOP} Python package, we constructed sparse, interpretable features indicating whether each OMOP concept (e.g., ``37003436'' for COVID vaccine, ``200219'' for abdominal pain) appeared at least once pre-index:
\[
x_{\text{OMOP},j} =
\begin{cases}
1 & \text{if count(concept } j) > 0 \\
  & \text{in } [\text{start},\text{index})\\
0 & \text{otherwise.}
\end{cases}
\]
This yields a high-dimensional ($>10^4$) binary vector $\textbf{x}_{\text{EHR-OMOP}}$ aligned to the pre-defined observation window. The concept IDs are drawn from the OMOP Common Data Model standardized vocabulary, which can be explored and referenced through the Athena concept database at \href{https://athena.ohdsi.org/}{Athena.OHDSI}.

\noindent\textbf{(ii) Structured-EHR foundation embedding (CEHR-GPT)}

We encoded each participant’s longitudinal pre-index EHRs into a fixed-length embedding using a structured-EHR foundation model, CEHR-GPT \citep{pang_cehr-gpt_2024}. It uses a novel patient representation that encodes complete visit timelines with demographic prompts, visit-type tokens, discharge information, and artificial time tokens, enabling transformer models to preserve temporal dependencies across heterogeneous OMOP domains when generating dense EHR embeddings $\textbf{x}_{\text{EHR-CEHR}}\in\mathbb{R}^{768}$.

\vspace{-3mm}
\subsubsection{\textbf{Wearable Data Representation}}
\textbf{(i) Wearable statistical summaries (221-dim)}

From daily Fitbit's physical activity, heart rate (including daily resting heart rate), and sleep (total duration and stage minutes: light, deep, REM), we computed per-channel summary statistics over an 180-day pre-index window with the most valid days (up to 1 year before prediction time): mean, standard deviation, minimum, maximum, and an ordinary-least-squares linear trend (slope over day index). Concatenating across primitives produced 221 features per participant:
$\textbf{x}_{\text{wear-sum}}\in\mathbb{R}^{221}$.
For each participant, wearable summaries were computed using an optimal 180-day window selected through a systematic search process: starting from the index date and moving backward in 7-day increments up to 365 days pre-index, all windows meeting the $\geq$ 50\% valid days threshold were evaluated, and the window with the largest number of valid days was selected for feature computation. No forward/backward filling was performed at the day level.

\vspace{5pt}
\noindent\textbf{(ii) Wearable time-series foundation embeddings (MOMENT-1-large)}

To capture temporal structure beyond summaries, we employed the time-series foundation model MOMENT-1-large \citep{goswami_moment_2024} to generate embeddings of wearable data. MOMENT-1-large is pretrained with a masked time-series modeling objective in a self-supervised manner and learns general representations from large-scale, multi-domain time-series data. In this study, we treated each participant’s multivariate daily sequences as a matrix input (same data window as the statistical summary feature $\textbf{x}_{\text{wear-sum}}$), with variables as channels and time as the sequence dimension. The model outputs a fixed-length sequence-level embedding of dimension
$\textbf{x}_{\text{wear-ts}} \in \mathbb{R}^{1024}$, 
which is used as the representation for downstream analysis.

\subsection{Feature Fusion Approaches}
\label{sec:fusion}
We evaluated six late- and feature-level fusion strategies combining EHR (OMOP or CEHR) and wearable (summary or time-series) representations. 
When using OMOP representations for EHR, since $\textbf{x}_{\text{EHR-OMOP}}$ has over 10$^4$ dimensions, we conducted top feature selection for both EHR and wearable embeddings before concatenating them (see top two rows in \tableref{tab:fusion_methods}).
When using CEHR representations, we experimented with wearable embedding concatenation ($\textbf{x}_{\text{EHR-CEHR}}\in \mathbb{R}^{1024}$ and $\textbf{x}_{\text{wear-ts}}\in \mathbb{R}^{768}$ have similar length), concatenation with substitution ($\textbf{x}_{\text{EHR-CEHR}}$ is significantly longer than $\textbf{x}_{\text{wear-sum}}\in \mathbb{R}^{221}$),  and weighting (applicable for both $\textbf{x}_{\text{wear-sum}}$ and $\textbf{x}_{\text{wear-ts}}$), as indicated in the bottom four rows in \tableref{tab:fusion_methods}.
All constituent features were standardized (z-score) on the training set prior to fusion; the learned scaler was reused on test.

\subsection{Evaluation}
\label{sec:evaluation}


After feature fusion across the two data sources, classification used $\ell_2$-regularized logistic regression. The inverse regularization strength $C$ was selected by internal cross-validation on the training set and then fixed for test evaluation. After standardization, any remaining NaNs were set to zero.

For each health outcome, all model selection or finetuning used the training set only. The held-out test set for each outcome, disjoint from external CEHR training, was used \emph{once} for final evaluation. To obtain robust \emph{paired} comparisons when evaluating a model trained on the test set, we generated 100 bootstrap resamples of the test set by sampling participants with replacement to the original test size. Resamples were prevalence-constrained to reflect the true positive rate with stratified sampling, requiring at least two positive samples. The \emph{same} resampled indices were applied across all methods in each bootstrap round for a paired t-test. AUROC was the primary endpoint, and we report both mean and standard deviations of the AUROC scores. For each outcome, we compared the best of the EHR-only baselines (CEHR-only, OMOP-only) and the best of the fusion models.

\subsection{Pipeline Extensibility}
Our pipeline is disease-agnostic: to study new health conditions, researchers need only specify relevant phenotyping codes (e.g., SNOMED, PheCodes, or medications) to define incident cases, after which the same cohort construction, feature extraction, and fusion steps can be applied directly. Our pipeline code was open-sourced at \href{https://anonymous.4open.science/r/AoU_Wearable_Valuation-07F3/README.md}{GITHUB/AoU\_Wearable}, and the All of Us dataset we used in this study can be accessed via institutional permission after ethics training and registration. 

\section{Results}

\subsection{Wearable Integration Leads to Consistent Performance Gain}

\begin{figure}[t]
  \centering
  \includegraphics[width=1.04\linewidth]{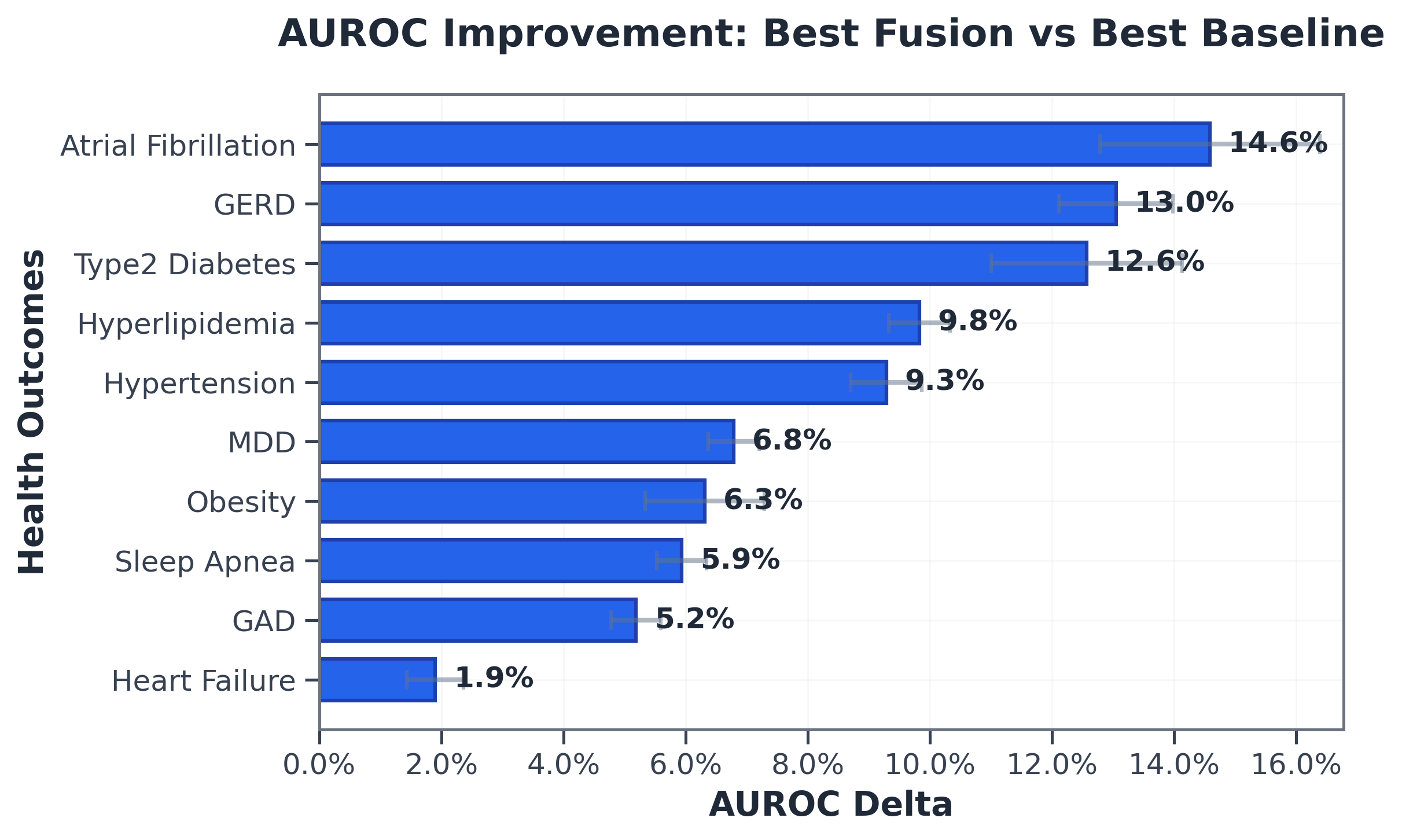}
  \vspace{-8mm}
  \caption{Performance gain of EHR-wearable multimodal models over the best EHR baseline. Error bars represent 95\% confidence intervals.}
  \label{fig:results_barplot}
  \vspace{-6mm}
\end{figure}

\figureref{fig:results_barplot} illustrates that for each of the 10 diseases, the best fusion model achieves a higher AUROC than the strongest EHR-only baseline, with gains ranging from modest improvements in Heart Failure (1.9\%) to substantial increases for metabolic and psychiatric outcomes such as Type II Diabetes (12.6\%), Hyperlipidemia (9.8\%), and Major Depressive Disorder (6.8\%). All improvements demonstrated statistical significance with $p<0.001$. This universal improvement in performance underscores the complementary value of wearable-derived signals when combined with EHR data, which provide additional physiological and behavioral information not captured in clinical records. 

\begin{figure*}[htbp]
  \centering
  \vspace{-4mm}
  \includegraphics[width=0.95\linewidth]{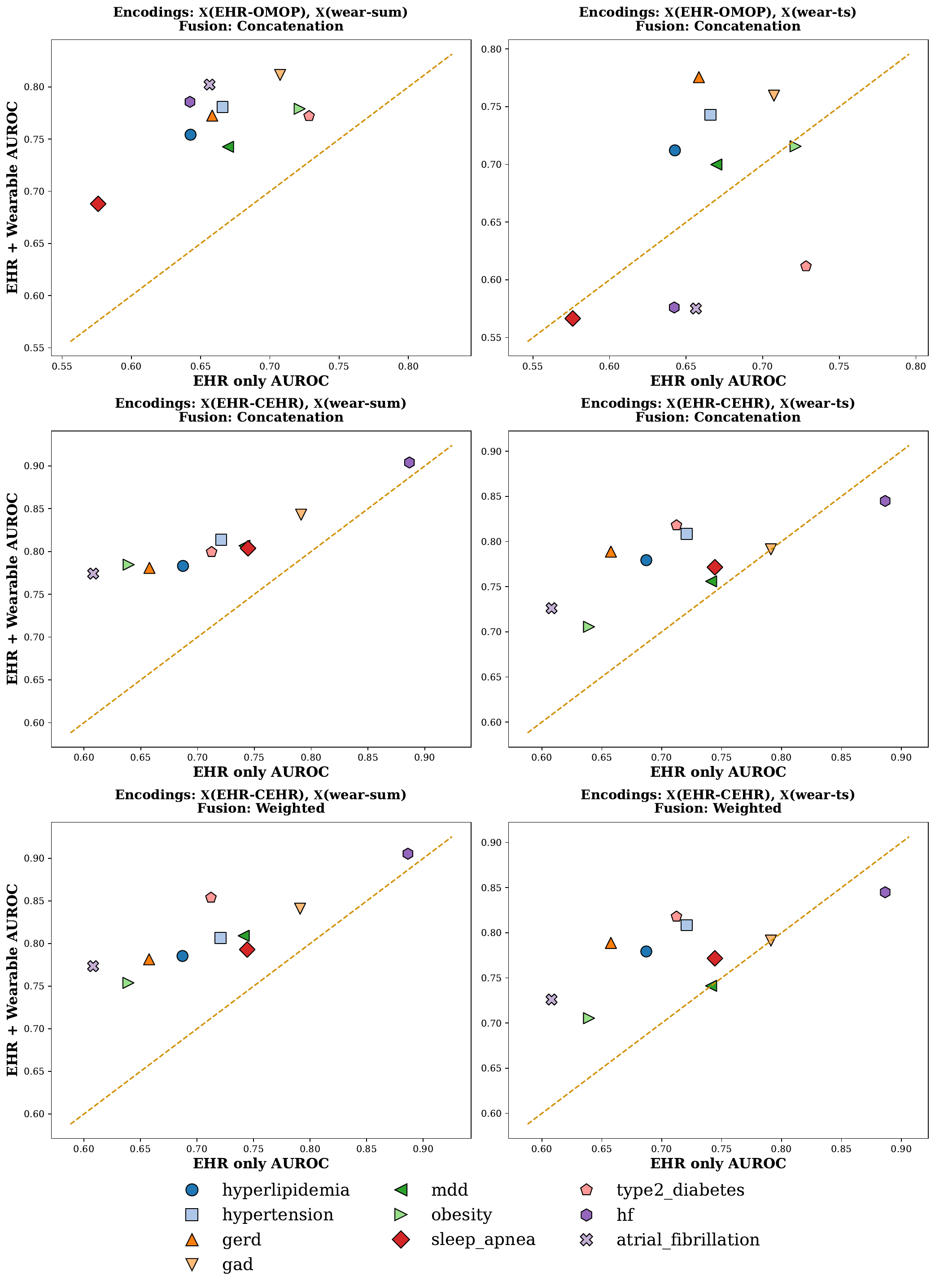}
  \vspace{-5mm}
  {\caption{Mean AUROC of EHR-only models vs EHR + wearables across fusion strategies and feature backbones. Each point is a clinical outcome (legend).}}
  \label{fig:results_panel}
\end{figure*}

In addition to the aggregated advantage, we further compare each pair of the multimodal model and the EHR baseline across all encoding methods and fusion strategies.
As illustrated in \figureref{fig:results_panel}, nearly every data point lies above the diagonal line of equality, indicating universal improvements when wearable features are incorporated into EHR-based prediction models.
Comparing each outcome's best EHR-wearable fusion model against its strongest EHR-only baseline yielded a mean AUROC improvement of $+8.5\%$ over 100 paired bootstrap resamples.

For additional details on the exact performances of each fusion strategy for each health outcome, please refer to Supplementary \tableref{tab:summary_results} and Supplementary \tableref{tab:all_results}.


\subsection{EHR Foundation Model Improves EHR-only Performance but Shows Less Benefit with Wearable Data}
We further compare the traditional feature engineering-based and modern foundation model-based encoding methods on EHR data.
Our analysis reveals distinct patterns in encoding effectiveness across different clinical contexts, as shown in  Supplementary \tableref{tab:all_results}.

The EHR foundation model embeddings $\textbf{x}_{\text{EHR-CEHR}}$ outperformed traditional OMOP encoding $\textbf{x}_{\text{EHR-OMOP}}$ in most EHR-only predictions (6 out of 10 health outcomes, with an overall mean AUROC improvement of $+5.2\%$), and its advantage became more consistent when integrated with wearable data (9 out of 10 outcomes, with an average improvement of $+4.7\%$ over OMOP-based fusion methods).

These results suggest that the rich, chronological patient representations learned by the foundation model provide effective substrates for fusion with continuous physiological signals from wearable devices.

\subsection{Manual Wearable Features Outperform the Time-series Foundation Model}
We follow a similar process to compare the two encoding methods for the wearable data.
Interestingly, across all 10 health outcomes, embeddings $\textbf{x}_{\text{wear-ts}}$ using the MOMENT time-series foundation model consistently underperform the summary features $\textbf{x}_{\text{wear-sum}}$, in both wearable-alone (average $\Delta$AUROC $-4.4\%$) and multimodal setups (average $\Delta$AUROC $-3.8\%$).
These results show that simple summary-level wearable features provide more meaningful information for health outcome prediction tasks, compared to the more sophisticated foundational model.
Our analysis suggests that such domain-agnostic time-series embedding methods are not ready to be applied directly for health-related tasks. We have more discussion about these insights in Sec.~\ref{discussion:future_fm}.

\section{Discussion}

\subsection{The Complementary Value of Wearable Data across Health Outcomes}
\label{discussion:wearable_value}
The integration of wearable data consistently elevated prediction performance across a diverse set of diseases.
The particularly strong performance of summary-level wearable features suggests that day-level aggregations of activity, heart rate, and sleep patterns provide stable and interpretable augmentation to clinical records. The success across such diverse conditions---from metabolic disorders like diabetes and obesity to mental health conditions like depression and anxiety---indicates that basic physiological monitoring captures fundamental health signals relevant to multiple disease processes.

Meanwhile, we also observe that the magnitude of this benefit varied by conditions. More specifically, for heart failure, the addition of wearable data showed only modest improvements ($\Delta$AUROC +1.9\%). This can come from several potential reasons. Other than the fact that it already began with a strong baseline performance (AUROC 0.886) with less room for improvement, heart failure, as an acute health condition, may show fewer signals captured through daily wearable devices.

Our comprehensive and systematic evaluation provides the first nuanced picture of the predictive value of wearable data for diverse health outcomes.

\subsection{Towards Multimodal Foundation Models with EHR and Wearables for Health}
\label{discussion:future_fm}
Our findings offer new insights into the current capabilities and limitations of foundation models for health prediction. The EHR foundation model, CEHR-GPT, showed an advantage in EHR-only settings, and this advantage was not only maintained but became more consistent when combined with wearable features.
The superior performance of CEHR-based multimodal models indicates that foundation model embeddings capture complementary information that synergizes well with wearable data, even when using relatively simple fusion methods like concatenation and weighted averaging.

Conversely, the general-purpose time-series foundation model, MOMENT, consistently underperformed compared to domain-specific feature engineering on wearable data. This performance gap is likely attributable to a domain mismatch; a model pre-trained on generic time-series data may not generate embeddings that capture the specific physiological patterns most relevant to long-term health outcomes, a phenomenon that is supported by some recent work in other domains~\citep{tan2024language,gu2025time}. Manually engineered features, such as average resting heart rate or daily step counts, are grounded in clinical knowledge and effectively act as powerful, low-noise summaries of behavior, which proves more effective for this specific predictive task than the abstract representations from a generalist model.

These results highlight that the development of EHR and wearable foundation models, especially when developed for health applications, should not proceed in isolation.
The path forward lies in creating multimodal foundation models that are pre-trained on integrated EHR and wearable data.
Recent research on multimodal foundation models (e.g., imaging with clinical notes~\citep{zhang2025multimodal}, imaging with EHR~\citep{liu2025metagp}) has started to explore this direction.
Such models could learn a unified patient representation and move beyond simply combining outputs from single-modality models and toward a holistic, dynamic understanding of patient health by including longitudinal and supplementary information between clinical visits.

\subsection{Limitations}
\label{discussion:limitation}
Our study has several limitations that open avenues for future research. First, our fusion strategies were limited to straightforward approaches; future work should investigate more sophisticated techniques, such as attention-based mechanisms or cross-modal transformers, to better model the interactions between EHR and wearable data streams.
Second, our analysis of foundation models was not exhaustive. A broader evaluation including other emerging EHR models (e.g., MOTOR~\citep{steinberg_motor_2023}) and wearable-specific time-series models (e.g., WBM~\citep{erturk2025beyond}, LSM~\citep{xu2025lsm}, SensorLM~\citep{zhang2025sensorlm}) is a critical next step, contingent on their availability as open-source resources.
Furthermore, the generalizability of our findings may be constrained by our specific study population and wearable device types, which, being composed of research volunteers, may exhibit a ``healthy user'' bias and not fully represent the broader, higher-risk patient population. Real-world deployment would also need to address irregular data patterns and varying patient compliance. Finally, the 180-day washout period used to define incident disease was selected based on precedent in existing literature and was not empirically examined or optimized for each of the ten distinct conditions. We aim to address these limitations through continued research and future publications.


\section{Conclusion}
In this work, we conducted the first large-scale, systematic evaluation of integrating consumer wearable data with EHRs for predicting ten diverse health outcomes. Our comprehensive benchmark demonstrates that augmenting episodic clinical data with continuous, real-world data from wearables yields substantial and consistent improvements in predictive accuracy, with an average AUROC increase of +8.5\%. Our analysis of different encoding strategies further revealed that while foundation models hold promise, domain-specific feature engineering remains highly effective for wearable data, and the utility of current single-modality EHR foundation models can be attenuated in a multimodal context. By providing robust evidence of the complementary value of wearable data and open-sourcing our pipeline, this study paves the way for developing more holistic and personalized predictive models in future AI health research.


\bibliography{jmlr-sample}

\clearpage
\appendix
\renewcommand{\tablename}{Supplementary Table}

\setcounter{table}{0} 

\renewcommand{\figurename}{Supplementary Figure}
\setcounter{figure}{0} 

\section{Cohort Definition Example}\label{apd:cohort_example}

We conducted a retrospective cohort study in the All of Us Registered Tier, restricting to participants with both OMOP-standardized EHRs and connected Fitbit data. All time references are anchored to a participant-specific prediction (``index'') time defined below. In this section we show an example process of cohort definition with specific numbers. 

\subsection{Source population and phenotyping}
From 8{,}477 All of Us participants with valid Fitbit and EHR linkage, we identified 2{,}148 participants with any evidence of major depressive disorder (MDD) using the \emph{union} of (i) OMOP concept IDs for MDD, (ii) PheCode mappings for depressive disorders, and (iii) depression-specific medications. Phenotyping was performed strictly prior to index.

\subsection{Incident case definition and index time}
Incident cases required the first MDD evidence to occur \emph{at least} 180 days after Fitbit tracking began, and evidence of EHR activity both before and after Fitbit start. For positives, the index time was set to one day prior to the first MDD flag, yielding 300 incident cases. For negatives, the index time was set to one day prior to the last observed EHR record.

\subsection{Control pool and analytic cohort}
The control pool comprised participants with (i) no MDD evidence at any time, (ii) $\geq$180 days of valid Fitbit data, and (iii) EHR activity both before and after Fitbit start, producing 2{,}701 negatives. These criteria formed an initial analytic cohort of 3{,}001 participants (11.4\% positives).

\subsection{Wearable/EHR quality filters and final dataset}
We applied pre-specified quality control (QC) on wearable data density/regularity and device sanity checks, and excluded participants failing minimal EHR coverage around index. After QC, 2{,}391 participants remained (12.6\% positives).

\subsection{Holdout protocol and leakage safeguards}
To avoid leakage when comparing to an external CEHR model trained on other AoU participants, we restricted the held-out test set to participants \emph{not} present in CEHR training (477 participants; 13.0\% positives) and used the remainder for model development (1{,}914; 10.8\% positives). All feature extraction windows (EHR and wearable) were strictly pre-index (no look-ahead).

\paragraph{Summary of counts.}
8{,}477 $\rightarrow$ 3{,}001 (phenotyping and design constraints) $\rightarrow$ 2{,}391 (data quality control) split into 1{,}914 development and 477 held-out test participants.

\begin{figure}[htbp]
\floatconts
  {fig:cohort_example}
  {\caption{Example cohort definition steps and numbers for Major Depressive Disorder}}
  {\includegraphics[width=0.80\linewidth]{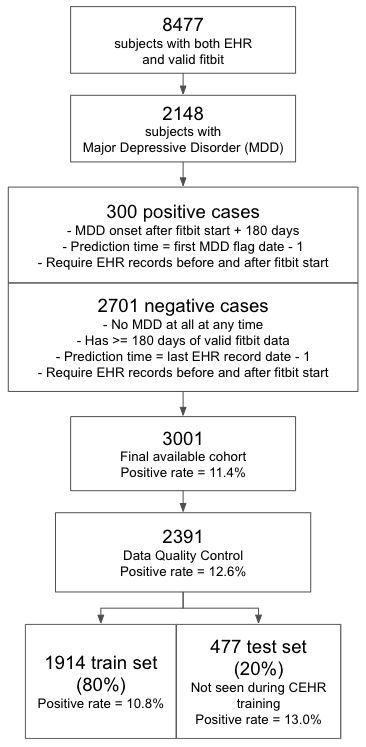}}
\end{figure}

\onecolumn
\section{Detailed Results}\label{apd:more_results}

Abbreviations used in the tables: To optimize space utilization in the performance comparison tables, we employ standardized abbreviations for health outcomes and methodological components. Health outcomes are abbreviated as follows: HLD (Hyperlipidemia), HTN (Hypertension), GERD (Gastroesophageal Reflux Disease), GAD (Generalized Anxiety Disorder), MDD (Major Depressive Disorder), OB (Obesity), SA (Sleep Apnea), T2D (Type 2 Diabetes), HF (Heart Failure), and AF (Atrial Fibrillation). Methodological abbreviations include CEHR (Clinical Element-based Health Records), OMOP (Observational Medical Outcomes Partnership), and TS Embed (Time Series Embedding). These abbreviations enable comprehensive presentation of performance metrics across all outcomes and fusion approaches while maintaining table readability and fitting within standard page constraints.

\begin{table*}[!htb]
\centering 
\setlength{\tabcolsep}{3pt} 
\Large
\renewcommand{\arraystretch}{1.4}
\caption{Performance (AUROC) of best feature fusion between EHR and wearable data across health outcomes against EHR baselines.}
\label{tab:summary_results}
\resizebox{\textwidth}{!}{
\begin{tabular}{l|cccccccccc}
\toprule 
& \textbf{HLD} & \textbf{HTN} & \textbf{GERD} & \textbf{GAD} & \textbf{MDD} & \textbf{OB} & \textbf{SA} & \textbf{T2D} & \textbf{HF} & \textbf{AF} \\
\midrule
OMOP & $0.643_{\pm 0.034}$ & $0.666_{\pm 0.040}$ & $0.658_{\pm 0.032}$ & $0.707_{\pm 0.036}$ & $0.670_{\pm 0.037}$ & $0.721_{\pm 0.035}$ & $0.576_{\pm 0.040}$ & $0.728_{\pm 0.060}$ & $0.642_{\pm 0.074}$ & $0.656_{\pm 0.074}$ \\
CEHR & $0.687_{\pm 0.037}$ & $0.721_{\pm 0.034}$ & $0.658_{\pm 0.039}$ & $0.791_{\pm 0.031}$ & $0.741_{\pm 0.033}$ & $0.639_{\pm 0.046}$ & $0.744_{\pm 0.037}$ & $0.712_{\pm 0.067}$ & $0.886_{\pm 0.033}$ & $0.608_{\pm 0.054}$ \\
Feature Fusion & 
$\bm{0.785_{\pm 0.033}}^{(4)}$ &
$\bm{0.814_{\pm 0.028}}^{(3)}$ &
$\bm{0.789_{\pm 0.030}}^{(5)}$ &
$\bm{0.843_{\pm 0.030}}^{(3)}$ &
$\bm{0.809_{\pm 0.027}}^{(4)}$ &
$\bm{0.785_{\pm 0.039}}^{(3)}$ &
$\bm{0.804_{\pm 0.034}}^{(3)}$ &
$\bm{0.854_{\pm 0.062}}^{(4)}$ &
$\bm{0.905_{\pm 0.035}}^{(4)}$ &
$\bm{0.802_{\pm 0.050}}^{(1)}$ \\
\midrule
Significance & *** & *** & *** & *** & *** & *** & *** & *** & *** & *** \\
\bottomrule
\end{tabular}
}
\end{table*}

\textbf{Notes:} The baseline uses two types of EHR encodings: OMOP and CEHR. Significance levels: *** p$<$0.001, ** p$<$0.01, * p$<$0.05. Best fusion methods for each outcome:  (1) [$\textbf{x}_{\text{EHR-OMOP}}$, $\textbf{x}_{\text{wear-sum}}$] (Concat); (3) [$\textbf{x}_{\text{EHR-CEHR}}$, $\textbf{x}_{\text{wear-sum}}$] (Concat); (4) [$\textbf{x}_{\text{EHR-CEHR}}$, $\textbf{x}_{\text{wear-ts}}$] (Weighted); (5) [$\textbf{x}_{\text{EHR-CEHR}}$, $\textbf{x}_{\text{wear-sum}}$] (Concat). See next table.

\begin{table*}[!htb]
\centering
\Large
\setlength{\tabcolsep}{3pt}
\renewcommand{\arraystretch}{1.4}
\caption{Performance comparison across all baseline models and feature fusion methods.}
\label{tab:all_results}
\resizebox{\textwidth}{!}{
\begin{tabular}{l|cccccccccc}
\toprule
\textbf{Encoding \& Fusion Methods} & \textbf{HLD} & \textbf{HTN} & \textbf{GERD} & \textbf{GAD} & \textbf{MDD} & \textbf{OB} & \textbf{SA} & \textbf{T2D} & \textbf{HF} & \textbf{AF} \\
\midrule
$\textbf{x}_{\text{EHR-OMOP}}$ & $0.643_{\pm 0.034}$ & $0.666_{\pm 0.040}$ & $0.658_{\pm 0.032}$ & $0.707_{\pm 0.036}$ & $0.670_{\pm 0.037}$ & $0.721_{\pm 0.035}$ & $0.576_{\pm 0.040}$ & $0.728_{\pm 0.060}$ & $0.642_{\pm 0.074}$ & $0.656_{\pm 0.074}$ \\
$\textbf{x}_{\text{EHR-CEHR}}$ & $0.687_{\pm 0.037}$ & $0.721_{\pm 0.034}$ & $0.658_{\pm 0.039}$ & $0.791_{\pm 0.031}$ & $0.741_{\pm 0.033}$ & $0.639_{\pm 0.046}$ & $0.744_{\pm 0.037}$ & $0.712_{\pm 0.067}$ & $0.886_{\pm 0.033}$ & $0.608_{\pm 0.054}$ \\
$\textbf{x}_{\text{wear-sum}}$ & $0.804_{\pm 0.029}$ & $0.807_{\pm 0.031}$ & $0.767_{\pm 0.033}$ & $0.833_{\pm 0.028}$ & $0.797_{\pm 0.026}$ & $0.762_{\pm 0.038}$ & $0.741_{\pm 0.036}$ & $0.790_{\pm 0.058}$ & $0.778_{\pm 0.059}$ & $0.791_{\pm 0.053}$ \\
$\textbf{x}_{\text{wear-ts}}$ & $0.753_{\pm 0.034}$ & $0.812_{\pm 0.026}$ & $0.792_{\pm 0.029}$ & $0.745_{\pm 0.030}$ & $0.671_{\pm 0.034}$ & $0.695_{\pm 0.041}$ & $0.687_{\pm 0.046}$ & $0.774_{\pm 0.062}$ & $0.742_{\pm 0.068}$ & $0.764_{\pm 0.058}$ \\
\midrule
(1) [$\textbf{x}_{\text{EHR-OMOP}}$, $\textbf{x}_{\text{wear-sum}}$] (Concat) & $0.754_{\pm 0.036}$ & $0.781_{\pm 0.029}$ & $0.773_{\pm 0.028}$ & $0.811_{\pm 0.029}$ & $0.743_{\pm 0.027}$ & $0.779_{\pm 0.038}$ & $0.688_{\pm 0.041}$ & $0.772_{\pm 0.065}$ & $0.786_{\pm 0.064}$ & $\bm{0.802_{\pm 0.050}}$ \\
(2) [$\textbf{x}_{\text{EHR-OMOP}}$, $\textbf{x}_{\text{wear-ts}}$] (Concat) & $0.712_{\pm 0.028}$ & $0.743_{\pm 0.029}$ & $0.776_{\pm 0.027}$ & $0.760_{\pm 0.035}$ & $0.700_{\pm 0.030}$ & $0.716_{\pm 0.035}$ & $0.567_{\pm 0.032}$ & $0.612_{\pm 0.058}$ & $0.576_{\pm 0.081}$ & $0.575_{\pm 0.099}$ \\
(3) [$\textbf{x}_{\text{EHR-CEHR}}$, $\textbf{x}_{\text{wear-sum}}$] (Concat) & $0.783_{\pm 0.033}$ & $\bm{0.814_{\pm 0.028}}$ & $0.781_{\pm 0.031}$ & $\bm{0.843_{\pm 0.030}}$ & $0.807_{\pm 0.028}$ & $\bm{0.785_{\pm 0.039}}$ & $\bm{0.804_{\pm 0.034}}$ & $0.799_{\pm 0.068}$ & $0.904_{\pm 0.036}$ & $0.774_{\pm 0.045}$ \\
(4) [$\textbf{x}_{\text{EHR-CEHR}}$, $\textbf{x}_{\text{wear-sum}}$] (Weighted) & $\bm{0.785_{\pm 0.033}}$ & $0.806_{\pm 0.029}$ & $0.782_{\pm 0.031}$ & $0.841_{\pm 0.030}$ & $\bm{0.809_{\pm 0.027}}$ & $0.754_{\pm 0.041}$ & $0.793_{\pm 0.034}$ & $\bm{0.854_{\pm 0.062}}$ & $\bm{0.905_{\pm 0.035}}$ & $0.773_{\pm 0.045}$ \\
(5) [$\textbf{x}_{\text{EHR-CEHR}}$, $\textbf{x}_{\text{wear-ts}}$] (Concat) & $0.779_{\pm 0.032}$ & $0.808_{\pm 0.027}$ & $\bm{0.789_{\pm 0.030}}$ & $0.791_{\pm 0.033}$ & $0.756_{\pm 0.035}$ & $0.706_{\pm 0.042}$ & $0.772_{\pm 0.031}$ & $0.818_{\pm 0.070}$ & $0.845_{\pm 0.051}$ & $0.726_{\pm 0.055}$ \\
(6) [$\textbf{x}_{\text{EHR-CEHR}}$, $\textbf{x}_{\text{wear-ts}}$] (Weighted) & $0.779_{\pm 0.032}$ & $0.808_{\pm 0.027}$ & $0.789_{\pm 0.030}$ & $0.791_{\pm 0.033}$ & $0.741_{\pm 0.033}$ & $0.705_{\pm 0.042}$ & $0.772_{\pm 0.031}$ & $0.818_{\pm 0.070}$ & $0.845_{\pm 0.051}$ & $0.726_{\pm 0.055}$ \\
\bottomrule
\end{tabular}
}
\end{table*}

As shown in Supplementary \tableref{tab:summary_results}, the integration of wearable data produced strong improvements for chronic cardiovascular and metabolic conditions. For hyperlipidemia, the [$\textbf{x}_{\text{EHR-CEHR}}$, $\textbf{x}_{\text{wear-sum}}$] (Weighted) approach achieved an AUROC of $0.785$, representing a $+0.098$ improvement over the CEHR-only baseline of $0.687$ ($p = 1.4 \times 10^{-61}$). Similarly, hypertension prediction improved from a CEHR baseline of $0.721$ to $0.814$ using [$\textbf{x}_{\text{EHR-CEHR}}$, $\textbf{x}_{\text{wear-sum}}$] (Concat), yielding a $+0.093$ improvement ($p = 2.4 \times 10^{-53}$). Obesity prediction demonstrated remarkable improvement through [$\textbf{x}_{\text{EHR-CEHR}}$, $\textbf{x}_{\text{wear-sum}}$] (Concat), rising from an OMOP baseline of $0.721$ to $0.785$ ($+0.063$, $p = 8.7 \times 10^{-23}$). Type II diabetes achieved the highest final AUROC of $0.854$ through [$\textbf{x}_{\text{EHR-CEHR}}$, $\textbf{x}_{\text{wear-sum}}$] (Weighted), improving from an OMOP baseline of $0.728$ ($+0.126$, $p = 4.6 \times 10^{-29}$).

Gastroesophageal reflux disease showed substantial improvement as well, with [$\textbf{x}_{\text{EHR-CEHR}}$, $\textbf{x}_{\text{wear-ts}}$] (Concat) increasing performance from a baseline OMOP AUROC of $0.658$ to $0.789$, representing a $+0.131$ gain ($p = 2.1 \times 10^{-48}$). 

Mental health conditions also benefited significantly from wearable integration. Generalized anxiety disorder prediction improved from a strong CEHR baseline of $0.791$ to $0.843$ using [$\textbf{x}_{\text{EHR-CEHR}}$, $\textbf{x}_{\text{wear-sum}}$] (Concat), achieving a $+0.052$ improvement ($p = 6.0 \times 10^{-45}$), while major depressive disorder showed similar gains from $0.741$ to $0.809$ with [$\textbf{x}_{\text{EHR-CEHR}}$, $\textbf{x}_{\text{wear-sum}}$] (Weighted) ($+0.068$, $p = 2.7 \times 10^{-54}$).

Sleep apnea showed consistent improvement, rising from a CEHR baseline of $0.744$ to $0.804$ using [$\textbf{x}_{\text{EHR-CEHR}}$, $\textbf{x}_{\text{wear-sum}}$] (Concat) ($+0.059$, $p = 1.0 \times 10^{-49}$). 

Even outcomes with already strong baseline performance showed meaningful improvements. Heart failure, which had the highest baseline CEHR performance at $0.886$, still achieved a statistically significant improvement to $0.905$ using [$\textbf{x}_{\text{EHR-CEHR}}$, $\textbf{x}_{\text{wear-sum}}$] (Weighted) ($+0.019$, $p = 1.1 \times 10^{-12}$). Atrial fibrillation demonstrated substantial improvement, rising from an OMOP baseline of $0.656$ to $0.802$ with [$\textbf{x}_{\text{EHR-OMOP}}$, $\textbf{x}_{\text{wear-sum}}$] (Concat), representing a $+0.146$ gain ($p = 2.1 \times 10^{-29}$).

\section{Demographics}\label{apd:demographics}
\vspace{-8mm}
\begin{table*}[!htb]
\centering 
\setlength{\tabcolsep}{8pt} 
\renewcommand{\arraystretch}{0.8}
\caption{Age and sex distribution across disease cohorts. Values shown as mean $\pm$ SD for age and percentages for sex. Cases indicate positive disease outcomes; Controls indicate negative outcomes.}
\label{tab:demographics_age_sex}
\vspace{-2mm}
\resizebox{\textwidth}{!}{
\begin{tabular}{l|cccccccccc}
\toprule 
& \textbf{HLD} & \textbf{HTN} & \textbf{GERD} & \textbf{GAD} & \textbf{MDD} & \textbf{OB} & \textbf{SA} & \textbf{T2D} & \textbf{HF} & \textbf{AF} \\
\midrule
\textbf{Sample Size} & & & & & & & & & & \\
\quad Cases & 385 & 296 & 324 & 329 & 269 & 269 & 257 & 106 & 94 & 68 \\
\quad Controls & 1,412 & 1,659 & 1,874 & 2,080 & 2,122 & 2,040 & 2,300 & 2,539 & 2,762 & 2,827 \\
\midrule
\textbf{Age (years)} & & & & & & & & & & \\
\quad Cases & $55.6_{\pm 12.9}$ & $57.8_{\pm 12.8}$ & $55.7_{\pm 14.2}$ & $49.1_{\pm 15.0}$ & $48.8_{\pm 15.8}$ & $51.5_{\pm 14.3}$ & $55.5_{\pm 13.9}$ & $57.9_{\pm 12.0}$ & $64.1_{\pm 11.3}$ & $65.6_{\pm 11.5}$ \\
\quad Controls & $51.7_{\pm 15.0}$ & $53.4_{\pm 15.1}$ & $57.1_{\pm 15.2}$ & $60.5_{\pm 14.2}$ & $59.9_{\pm 14.4}$ & $58.3_{\pm 15.2}$ & $57.4_{\pm 15.3}$ & $57.4_{\pm 15.2}$ & $57.6_{\pm 14.8}$ & $57.8_{\pm 14.8}$ \\
\midrule
\textbf{Female (\%)} & & & & & & & & & & \\
\quad Cases & 73.5 & 65.5 & 68.8 & 76.6 & 75.1 & 75.5 & 64.6 & 65.1 & 60.6 & 54.4 \\
\quad Controls & 75.6 & 74.7 & 68.6 & 65.2 & 66.1 & 68.8 & 72.1 & 70.7 & 69.6 & 69.6 \\
\midrule
\textbf{Male (\%)} & & & & & & & & & & \\
\quad Cases & 24.7 & 33.1 & 29.9 & 20.7 & 21.9 & 22.3 & 33.5 & 32.1 & 38.3 & 44.1 \\
\quad Controls & 21.2 & 22.2 & 28.9 & 33.3 & 32.3 & 28.9 & 25.7 & 26.9 & 28.0 & 27.9 \\
\bottomrule
\end{tabular}
}
\end{table*}

\vspace{-8mm}
\begin{table*}[!htb]
\centering 
\setlength{\tabcolsep}{11pt} 
\renewcommand{\arraystretch}{0.8}
\caption{Race and ethnicity distribution across disease cohorts. Values shown as percentages of overall cohort (cases + controls combined).}
\label{tab:demographics_race_ethnicity}
\vspace{-2mm}
\resizebox{\textwidth}{!}{
\begin{tabular}{l|cccccccccc}
\toprule 
& \textbf{HLD} & \textbf{HTN} & \textbf{GERD} & \textbf{GAD} & \textbf{MDD} & \textbf{OB} & \textbf{SA} & \textbf{T2D} & \textbf{HF} & \textbf{AF} \\
\midrule
\textbf{Race (\%)} & & & & & & & & & & \\
\quad White & 84.9 & 86.2 & 85.3 & 85.9 & 86.1 & 86.3 & 85.8 & 86.5 & 85.7 & 85.6 \\
\quad Black/African American & 3.5 & 2.3 & 3.2 & 3.7 & 3.5 & 2.7 & 3.4 & 3.1 & 3.5 & 3.6 \\
\quad Asian & 2.6 & 3.2 & 3.1 & 2.8 & 3.0 & 2.9 & 2.6 & 2.4 & 2.7 & 2.7 \\
\quad Multiple/Other & 9.0 & 8.3 & 8.4 & 7.6 & 7.4 & 8.1 & 8.2 & 8.0 & 8.1 & 8.1 \\
\midrule
\textbf{Ethnicity (\%)} & & & & & & & & & & \\
\quad Not Hispanic/Latino & 93.0 & 93.1 & 93.6 & 93.4 & 93.5 & 93.2 & 93.4 & 93.4 & 93.5 & 93.5 \\
\quad Hispanic/Latino & 5.6 & 5.5 & 4.6 & 5.0 & 4.8 & 5.1 & 5.0 & 4.8 & 4.8 & 4.7 \\
\quad Prefer not to answer/Skip & 1.4 & 1.4 & 1.8 & 1.6 & 1.7 & 1.7 & 1.6 & 1.8 & 1.7 & 1.8 \\
\bottomrule
\end{tabular}
}

\raggedright
\small
\textbf{Disease abbreviations:} HLD = Hyperlipidemia; HTN = Hypertension; GERD = Gastroesophageal Reflux Disease; GAD = Generalized Anxiety Disorder; MDD = Major Depressive Disorder; OB = Obesity; SA = Sleep Apnea; T2D = Type 2 Diabetes; HF = Heart Failure; AF = Atrial Fibrillation.

\textbf{Note:} Multiple/Other race category includes individuals identifying as more than one population, other single populations, none of the listed categories, none indicated, and those who preferred not to answer.
\end{table*}

\onecolumn
\section{Wearable Feature Importance}\label{apd:feature_importance}

\vspace{-8mm}

\begin{table*}[!htb]
\centering 
\setlength{\tabcolsep}{8pt} 
\renewcommand{\arraystretch}{0.9}
\caption{Most frequent wearable features appearing in top 20 predictors across 10 health conditions. Checkmarks indicate feature appeared in top 20 for that condition.}
\label{tab:wearable_feature_importance}
\resizebox{\textwidth}{!}{
\begin{tabular}{l|cccccccccc|c}
\toprule 
\textbf{Feature Category} & \textbf{HLD} & \textbf{HTN} & \textbf{GERD} & \textbf{GAD} & \textbf{MDD} & \textbf{OB} & \textbf{SA} & \textbf{T2D} & \textbf{HF} & \textbf{AF} & \textbf{Freq.} \\
\midrule
\textbf{Heart Rate - Fat Burn Zone} & & & & & & & & & & & \\
\quad Min HR (mean, max, min) & \checkmark & \checkmark & \checkmark & \checkmark & \checkmark & \checkmark & \checkmark & \checkmark & \checkmark & \checkmark & 10/10 \\
\quad Zone ratio (max, std, mean) & \checkmark & \checkmark & \checkmark & \checkmark & -- & \checkmark & \checkmark & \checkmark & \checkmark & \checkmark & 9/10 \\
\quad Minutes in zone (max, std, mean) & \checkmark & \checkmark & \checkmark & \checkmark & -- & \checkmark & -- & \checkmark & \checkmark & \checkmark & 8/10 \\
\midrule
\textbf{Resting Heart Rate} & & & & & & & & & & & \\
\quad HR variability - std & \checkmark & \checkmark & -- & -- & -- & \checkmark & \checkmark & \checkmark & -- & -- & 5/10 \\
\quad Max resting HR & -- & -- & \checkmark & \checkmark & -- & -- & \checkmark & -- & \checkmark & -- & 4/10 \\
\quad Non-resting HR metrics & -- & \checkmark & -- & -- & \checkmark & \checkmark & \checkmark & -- & -- & -- & 4/10 \\
\midrule
\textbf{Sleep Metrics} & & & & & & & & & & & \\
\quad Total sleep time (mean, max) & \checkmark & -- & \checkmark & \checkmark & -- & -- & -- & \checkmark & -- & -- & 4/10 \\
\quad Wake duration (max, std) & -- & -- & -- & \checkmark & \checkmark & -- & -- & -- & -- & -- & 2/10 \\
\quad Sleep stage ratios & -- & -- & \checkmark & \checkmark & \checkmark & -- & \checkmark & -- & -- & -- & 4/10 \\
\quad Time in bed (mean, max) & \checkmark & -- & \checkmark & -- & -- & -- & -- & \checkmark & -- & -- & 3/10 \\
\midrule
\textbf{Activity \& Steps} & & & & & & & & & & & \\
\quad Daily steps (mean, std, min, trend) & -- & \checkmark & \checkmark & \checkmark & \checkmark & \checkmark & -- & -- & -- & -- & 5/10 \\
\quad Valid recording days & -- & \checkmark & \checkmark & -- & -- & \checkmark & \checkmark & -- & -- & -- & 4/10 \\
\quad Active minutes (light/fair/very) & -- & \checkmark & -- & -- & -- & \checkmark & \checkmark & \checkmark & -- & -- & 4/10 \\
\midrule
\textbf{Heart Rate - Cardio Zone} & & & & & & & & & & & \\
\quad Zone ratio \& minutes & \checkmark & -- & -- & -- & -- & \checkmark & -- & \checkmark & \checkmark & \checkmark & 5/10 \\
\quad Min/max HR in cardio & \checkmark & \checkmark & \checkmark & \checkmark & -- & \checkmark & -- & -- & \checkmark & \checkmark & 7/10 \\
\bottomrule
\end{tabular}
}
\raggedright
\small
\textbf{Disease abbreviations:} HLD = Hyperlipidemia; HTN = Hypertension; GERD = Gastroesophageal Reflux Disease; GAD = Generalized Anxiety Disorder; MDD = Major Depressive Disorder; OB = Obesity; SA = Sleep Apnea; T2D = Type 2 Diabetes; HF = Heart Failure; AF = Atrial Fibrillation.

\textbf{Note:} Features shown are aggregated categories where similar metrics (mean, max, min, std) are grouped together. Checkmark indicates at least one variant of the feature appeared in the top 20 most important predictors for that condition. Frequency column shows number of conditions where the feature appeared.
\end{table*}

The pattern of feature importance across conditions reveals clinically meaningful insights into disease-specific physiological signatures. Most notably, fat burn zone heart rate metrics (minimum heart rate and zone ratios) emerged as universal predictors across all 10 conditions, suggesting that cardiovascular efficiency during moderate-intensity activity serves as a fundamental indicator of overall health status. In contrast, disease-specific patterns highlighted distinct pathophysiological mechanisms: purely cardiac conditions (heart failure and atrial fibrillation) were predicted exclusively by heart rate-based features with no contribution from activity or sleep metrics, reflecting their primary dependence on cardiac function rather than lifestyle factors. Mental health conditions (GAD and MDD) showed unique importance of sleep disruption metrics—particularly wake duration and sleep stage architecture—which were largely absent in other conditions, aligning with established bidirectional relationships between sleep disturbances and psychiatric disorders. Interestingly, GERD shared this sleep signature with mental health conditions, consistent with nocturnal reflux symptoms disrupting sleep architecture. Physical activity metrics (daily steps, active minutes) emerged as important predictors for lifestyle-modifiable conditions (hypertension, obesity, GERD, sleep apnea) but were notably absent in the top features for purely cardiac conditions, suggesting that wearable-derived activity patterns may be particularly valuable for preventive screening and behavioral intervention targeting in metabolic and lifestyle-related diseases.

\end{document}